\documentstyle[aps,psfig,preprint]{revtex}

\newcommand{\beq}{\begin{equation}}
\newcommand{\eeq}{\end{equation}}
\newcommand{\beqa}{\begin{eqnarray}}
\newcommand{\eeqa}{\end{eqnarray}}

\newcommand{\ie}{i.e.}
\newcommand{\eg}{e.g.}

\newcommand{\etal}{{\it et al.}}

\newcommand{\ecm}{$\sqrt{s} = 10.58$ GeV}
\renewcommand{\b}{{\rm b}}
\renewcommand{\c}{{\rm c}}
\renewcommand{\d}{{\rm d}}
\newcommand{\e}{{\rm e}}
\newcommand{\g}{{\rm g}}
\newcommand{\p}{{\rm p}}
\newcommand{\q}{{\rm q}}
\newcommand{\s}{{\rm s}}
\renewcommand{\u}{{\rm u}}
\newcommand{\B}{{\rm B}}
\newcommand{\J}{{\rm J}}
\newcommand{\Q}{{\rm Q}}
\newcommand{\Z}{{\rm Z}}
\newcommand{\ee}{\e^+ \e^-}
\newcommand{\bbar}{\bar{\b}}
\newcommand{\cbar}{\bar{\c}}
\newcommand{\qbar}{\bar{\q}}
\newcommand{\Qbar}{\bar{\Q}}
\newcommand{\JP}{\J/\psi}
\newcommand{\Pp}{\psi'}
\newcommand{\mc}{m_{\rm c}}
\newcommand{\LQCD}{\Lambda_{{\rm QCD}}}

\newcommand{\PL}[3]{Phys.\ Lett.\ {\bf {#1}}, {#2} ({#3})}

\newcommand{\PRD}[3]{Phys.\ Rev.\ D {\bf {#1}}, {#2} ({#3})}
\newcommand{\PRL}[3]{Phys.\ Rev.\ Lett.\ {\bf {#1}}, {#2} ({#3})}
\newcommand{\ZPC}[3]{Z. Phys.\ {\bf C{#1}}, {#2} ({#3})}

\newcommand{\IJMPHA}[3]{Int.\ J.\ Mod.\ Phys. {\bf A{#1}}, {#2} ({#3})}

\begin{document}
\draft

\preprint{CERN--TH/97--261; \ NORDITA-97/60 P; \ TUM/T39-97-25}
\title{$\chi_{\c J}$ production in $\ee$ annihilation}
\author{G. A. Schuler\cite{GSbyline}}
\address{Theory Division, CERN, CH-1211 Geneva 23, Switzerland}
\author{M.\ V\"anttinen\cite{MVbyline}}
\address{NORDITA, Blegdamsvej 17, DK-2100 Copenhagen \O}
\date{\today}
\maketitle

\begin{abstract}
Inclusive production of $\chi_{\c J}$ in $\ee$ annihilation 
is an excellent probe of the role of higher Fock states in the
production of heavy quarkonia. Within the non-relativistic QCD
approach, contributions from the short-distance production of
colour-octet $\c\cbar$ pairs are significantly larger than
those from colour-singlet production. At the same time, 
$\chi_{\c J}$ production rates 
are significantly smaller than expected in the colour
evaporation approach. Measurements of $\chi_{\c J}$ production
at CLEO and future B-factories will thus constitute a major
test of theoretical approaches to the production of heavy quarkonia.
\end{abstract}


Calculations of heavy-quark production in $\ee$ collisions are now 
approaching a level of high accuracy. Fifteen years after the calculation 
of the next-to-leading-order (NLO) perturbative QCD corrections to the 
total {\em open} heavy-quark cross section \cite{JLZ82}, we have now seen 
the completion of the NLO corrections to three-jet cross sections with 
massive quarks \cite{BBU97}. Despite continuous efforts over nearly 
20 years, such a precision has not yet been reached in the calculation 
of cross sections of heavy {\em bound states} 
\cite{CEM,Keung81,BCY93,DKM94,CL96,Cho96,BC96,CKY96,Schuler97,BKLS97}.
Predictive power depends on the understanding of the long-distance
bound-state formation process, which is therefore at the heart of
current quarkonium physics. In this letter
we investigate the inclusive production of $P$-wave (orbital angular
momentum $L=1$) quarkonium states in $\ee$ collisions. This is the first
study of these processes since essentially \cite{estimate-footnote} all 
previous work was concerned with the $S$-wave $\JP$ or $\Upsilon$ mesons. 

For a sufficiently large quark mass $m$, a quarkonium bound state of
a heavy quark $\Q$ and its antiquark $\Qbar$ is a non-relativistic system. 
The spectroscopy of both charmonia and bottomonia is well described 
in non-relativistic potential models, where the quarkonia are considered 
to be pure $\Q\Qbar$ states bound by an instantaneous potential. 
The question arises of whether the production of heavy quarkonia is 
also dominated by this leading Fock state.
Is the $\Q\Qbar$ pair produced, already at short distances, in a
configuration that corresponds to the asymptotic 
valence Fock component of the quarkonium? If it is,
then all non-perturbative information 
in the theoretical prediction reduces to a single number, namely 
the coordinate-space wave function at the origin, which can be 
extracted, \eg\ from a potential-model calculation. 

It is nowadays widely believed that higher Fock states are of
decisive importance. Yet there exist differing approaches
when it comes to relating observable cross sections to short-distance
$\Q\Qbar$ production amplitudes. In the colour evaporation model 
(CEM) \cite{CEM}, the quarkonium production cross section is prescribed 
to be a (process-independent) fraction of the $\Q\Qbar$ production
cross section, below the physical threshold for the production of a pair
of heavy-light mesons. In the
non-relativistic QCD (NRQCD) approach \cite{NRQCD}, the 
transition from $\Q\Qbar$
to quarkonium is represented in terms of a multipole
expansion, which leads to scaling rules for transition probabilities
in terms of $v$, the mean heavy-quark velocity in the meson. 
Any given cross section is a linear combination of several
non-perturbative factors, multiplied by process-dependent hard
factors. To leading order in $\LQCD/\mu$ ($\mu \gtrsim mv$) the former
are well-defined process-independent NRQCD matrix elements (MEs). 

We have calculated the $\ee \rightarrow \chi_{\c J} + X$ cross
sections within the NRQCD approach, using values of non-perturbative
transition probabilities determined from measurements of other reactions.
On the one hand, the results strongly violate the process-independence 
of cross-section ratios assumed in the CEM. 
Thus already an experimental upper limit on $\chi_{\c J}$ production 
in $\ee$ annihilation can establish that non-perturbative effects 
play a decisive role in bound-state formation. On the other hand, 
the NRQCD results are dominated by colour-octet production channels. 
Observation of the predicted shapes and normalizations of the 
cross sections will therefore provide striking evidence of the importance 
of higher Fock states and the scaling of their contributions with $v$.

In NRQCD, any cross section is given as a series expansion in both
$\alpha_s(m)$ and $v$, to leading order in $\LQCD/\mu$. Let us first
consider the case of $S$-wave quarkonia, in particular the
$\JP$ ($J^{PC}=1^{--}$). 
The leading contribution in $v$ is 
given by processes
where the $\c\cbar$ pair is produced in the leading, colour-singlet
Fock state \cite{notation-footnote}, $|\c\cbar_1(^3S_1)\rangle$:
\beqa
 \ee & \rightarrow & \c\cbar_1(^3S_1) + \c\cbar \ ,
\label{cc13S1cc}
\\
 \ee & \rightarrow & \c\cbar_1(^3S_1) + \g\g
\ .
\label{cc13S1gg}
\eeqa
The order $v^2$ correction to $\JP$ production still involves only
the leading Fock state and can be considered as a relativistic correction
to the wave function. Subleading Fock states first contribute at relative
order $v^4$, where the short-distance production of a colour-octet
$\c\cbar$ pair is followed by dipole transitions to the final
quarkonium. The chromo-electric (E1) and chromo-magnetic (M1) 
dipole transition scale as $v^2$ (E1) and $v^4$ (M1), 
at least for $\alpha_s(\mc v)/\pi \ll 1$ \cite{Schuler97}.
Dominant colour-octet $\JP$ production processes at $\ee$ colliders are
\beqa
 \ee & \rightarrow & \c\cbar_8(^1S_0) + \g \ ,
\label{cc81S0g}
\\
 \ee & \rightarrow & \c\cbar_8(^3P_J) + \g \ ,
\label{cc83PJg}
\\
 \ee & \rightarrow & \c\cbar_8(^3S_1) + \q\qbar
         \hspace{5mm} (\q=\u,\d,\s)
\ .
\label{cc83S1qq}
\eeqa
Processes (\ref{cc81S0g},\ref{cc83PJg}) are enhanced by
$1/\alpha_s(\mc)$ relative to 
(\ref{cc13S1cc}), (\ref{cc13S1gg}),
and (\ref{cc83S1qq}) is enhanced, at large energies, by a
logarithm $\ln(s/\mc^2)$ with respect to (\ref{cc13S1cc}) (which, in turn,
dominates (\ref{cc13S1gg}) by a power $s/\mc^2$ at large $s$). 
Current estimates of the long-distance MEs 
$\langle {\cal O}_{1,8}^{\JP}(^{2S+1}L_J)\rangle$ \cite{NRQCD}, 
which parametrize the transition from a $\c\cbar_{1,8}(^{2S+1}L_J)$
state to the $\JP$, are listed, 
for example, in \cite{Schuler97}. At the centre-of-mass
energy studied by the CLEO experiment at the Cornell Electron Storage Ring,
$\sqrt{s} = 10.6$ GeV, the cross sections for the colour-singlet 
processes (\ref{cc13S1cc}) and (\ref{cc13S1gg}) are $0.20\,$pb and 
$0.35\,$pb, respectively, about one third of the experimental
cross section
$\sigma = 1.65 \pm 0.25\pm 0.33\,$pb \cite{Wolf}. 
Hence there is room for colour-octet contributions, which we estimate as 
$0.50\,$pb for (\ref{cc81S0g}), 
$0.60\,$pb for (\ref{cc83PJg}), and $0.01\,$pb for (\ref{cc83S1qq}). 

Conclusions on the presence of colour-octet mechanisms would, 
however, be premature since estimates of the total $\JP$ cross section 
suffer from large uncertainties. 
First, the colour-singlet MEs are known to $50\%$ at best, and a factor 
of $2$ uncertainty is certainly not exaggerated for the colour-octet MEs. 
Secondly, there could be large perturbative
corrections. And finally, truly relativistic corrections 
of order $v^2$ may also be large, cf.\ direct $\JP$ production 
in hadronic collisions \cite{CERN7170}. 

Therefore it has been suggested \cite{BC96} to study the energy distribution 
$\d\sigma /\d z$ of the $\JP$, where $z = 2 E_{\JP}/\sqrt{s}$. 
The predicted large colour-octet processes (\ref{cc81S0g}) 
and (\ref{cc83PJg}) are concentrated near $z=1$ and 
should thus dominate the upper end point of the $z$ spectrum. A 
dramatic change should be visible, notably in the polar angular 
distribution of the $\JP$ w.r.t.\ the beam axis. Unfortunately, 
measurements at large $z$ are plagued by a large background 
from radiative $\Pp$ production \cite{Wolf}.

Another potentially useful observable is the $\JP$ polarization,
which is measurable via the angular distributions of its decay leptons. 
Partial calculations already exist \cite{DKM94,BKLS97}, but the measurement
is definitely not easy.

We propose the study of $\chi_{\c J}$ production as a means 
of investigating the importance of colour-octet contributions 
in quarkonium production. Colour-octet mechanisms will
leave distinct footprints in the total cross section, and in the
energy and polar-angle distributions.
%
Contrary to $S$-wave quarkonia, the $J^{++}$ $\chi_{\c J}$ mesons
receive contributions from one of their higher Fock states
$|\c\cbar_8(^3S_1)\g\rangle$ already at leading order in $v$:
the associated E1 transition suppresses the cross section by
$v^2$, but this suppression is compensated because an $S$-wave
operator scales as $1/v^2$ relative to a $P$-wave operator.

Colour-singlet production of $\chi_{\c J}$ proceeds through 
\beq
  \ee \rightarrow  \c\cbar_1(^3P_J) + \c\cbar
\ .
\label{cc13PJcc}
\eeq
Note that the contribution from 
\beq
  \ee  \rightarrow  \c\cbar_1(^3P_J) + \g\g
\label{cc13PJgg}
\eeq
is zero for pure photon exchange in the $s$ channel. Also the
$\Z$-exchange contribution is negligible, at low energies because
it is proportional to $s/M_{\Z}^2$, and at high energies because
(\ref{cc13PJgg}) is suppressed by $\mc^2/s$ relative to (\ref{cc13PJcc}).

Colour-octet contributions to the energy distribution of $\chi_{\c J}$ 
away from $z=1$ arise from (\ref{cc83S1qq}) and the processes
\beqa
 \ee  & \rightarrow &  \c\cbar_8(^3S_1) + \c\cbar
\ ,
\label{cc83S1cc} \\
 \ee & \rightarrow & \c\cbar_8(^3S_1) + \g\g
\ .
\label{cc83S1gg}
\eeqa
All three processes possess the same scaling in $v$ and $\alpha_s(\mc)$ 
as the colour-singlet process (\ref{cc13PJcc}). 
Two further colour-octet processes dominate near the $z=1$ end-point,
namely (\ref{cc81S0g}) and (\ref{cc83PJg}). The process 
\beq
 \ee \rightarrow \c\cbar_8(^3S_1) + \g
\label{cc83S1g}
\eeq 
is negligible for reasons similar to (\ref{cc13PJgg}). 

The calculation of processes 
(\ref{cc83S1qq},\ref{cc13PJcc},\ref{cc83S1cc},\ref{cc83S1gg})
is standard but tedious. We calculate the distributions in both 
the energy and the polar angle $\theta$ of the $\chi_{\c J}$,
\beq
  \frac{\d^2 \sigma}{\d z\, \d \cos\theta} = S(z)\, \left[
  1 + \alpha(z) \right]
\ .
\label{Salphadef}
\eeq
Details will be presented elsewhere. Here we comment on the relation
of our results and previous calculations.

At high energies, $\mc^2/s \rightarrow 0$, 
the cross section for (\ref{cc13PJcc}) reduces to 
\beq
  \frac{\d \sigma}{\d z} = 2\, \sigma(\ee \to \c\cbar)\, 
  \frac{1}{\mc^5}\, \langle {\cal O}_1^{\chi_{\c J}} ( ^3P_J) \rangle\, 
  D^{(1)}_{\c \rightarrow \c\cbar(^3P_J)}(z) \ ,
\eeq
where $D^{(1)}_{\c \rightarrow \c\cbar(^3P_J)}$ are the partonic 
colour-singlet charm fragmentation functions for 
$\c \rightarrow \c\cbar_1(^3P_J) + \c$.  
Our expressions for these functions agree with \cite{Yuan94,Ma96}. 

The fragmentation limit of (\ref{cc83S1cc}) is more involved. 
Writing it as 
\beq
 \frac{\d \sigma}{\d z} =  2\, \sigma(\ee \to \c\cbar)\, 
     \frac{1}{\mc^3}\,   \langle {\cal O}_8^{\chi_{\c J}}(^3S_1) \rangle\, 
  \hat{D}_8(z) \ ,
\label{hatD8}
\eeq
we observe that $\hat{D}_8(z)$ agrees with (3.6) in \cite{Ma96} 
if we take $\mu^2 = z^2\, (1-z)\, s$ in the result of \cite{Ma96}.
Note that (\ref{hatD8}) is not the sum of two fragmentation processes,
$\ee \rightarrow \c\cbar$ followed by $\c \rightarrow \chi_J$ or 
$\cbar \rightarrow \chi_J$ and $\ee \rightarrow \c\cbar \g$ followed
by $\g \rightarrow \chi_J$, as is the high-energy limit of
(\ref{cc83S1qq}). We can decompose (\ref{hatD8}) as
\beqa
  \frac{\d\sigma}{\d z} & = &
  \frac{1}{\mc^3}\, \langle O_8^{\chi_J}(^3S_1) \rangle \biggl\{ 
  2\, \sigma_{\c\cbar}\, \left[ 
  D^{(8)}_{\c \rightarrow \c\cbar(^3S_1)}(z)
  + R_8(z) \right] 
  \nonumber \\
  & & \left. \mbox{} + \int_z^1\, \frac{\d \hat{z}}{\hat{z}}\, \left.
  \frac{\d \sigma_{\c\cbar \g}}{\d y}\right|_{y=z/\hat{z}}\, 
     D^{(8)}_{\g \rightarrow \c\cbar(^3S_1)}(\hat{z}) \right\}
\ ,
\label{Dhat8}
\eeqa
where $\sigma_{\c\cbar, \c\cbar\g}$ is the cross section for 
$\ee \rightarrow \c\cbar, \c\cbar \g$; the gluon fragmentation
function is simply
$D^{(8)}_{\g \rightarrow \c\cbar(^3S_1)}(z) = \pi \alpha_s \delta(1-z)/24$, 
and $D^{(8)}_{\c \rightarrow \c\cbar(^3S_1)}$
is obtained by appropriately removing the NRQCD ME and changing the colour
factor in the colour-singlet $\c \rightarrow \JP$ fragmentation function
given in \cite{BCY93}. Also, $D^{(8)}_{\c \rightarrow \c\cbar(^3S_1)}$
coincides with the equal-flavour limit of the
$\bbar \rightarrow \bbar\c_8(^3S_1) + \cbar$ fragmentation function of
\cite{Yuan94}. However, from (\ref{Dhat8}) it is clear that
$D^{(8)}_{\c \rightarrow \c\cbar(^3S_1)}$ is not the colour-octet part
of the $\c \rightarrow \chi_{\c J}$ fragmentation function, as conjectured
in \cite{Yuan94}. Interference terms present in the
$\c\cbar_8(^3S_1) + \c\cbar$ diagrams for the equal-flavour case
(the non-zero remainder term $R_8$) forbid a simple rescaling of the
colour-singlet result. 

Our cross section for (\ref{cc83S1gg}) agrees with \cite{CL96} if we
appropriately replace the colour-singlet NRQCD ME and overall colour
factors by their colour-octet counterparts. Finally, our cross section for
(\ref{cc83S1qq}) reduces to a form that is analogous to the
$\theta$-integrated expression given in \cite{CKY96}. 

For our numerical results we use $\mc = 1.48\,$GeV, $\alpha_s = 0.28$,  
$\alpha_{{\rm em}} = 0.0075$, and values for the NRQCD MEs as given 
in \cite{Schuler97}. Two MEs have not yet been determined from any
experiment. We estimate their magnitude by scaling other known MEs:
\beqa
  \langle {\cal O}_8^{\chi_{\c J}}(^1S_0) \rangle
  & \approx & v^4\,  \frac{1}{3}\, 
              \langle {\cal O}_8^{\chi_{\c J}}(^3S_1) \rangle 
  \nonumber \\
  & = & \frac{2 J+1}{3}\, 2.8 \times 10^{-4}\, {\rm GeV}^3
\ ,
\label{element1S0} 
\\
  \langle {\cal O}_8^{\chi_{\c J}}(^3P_0) \rangle
  & \approx & \frac{2J+1}{3} \,   
              \frac { \langle {\cal O}_8^{\JP}(^3S_1) \rangle}
              { \langle {\cal O}_1^{\JP}(^3S_1) \rangle} \,
              \langle {\cal O}_1^{\chi_{\c 1}}(^3P_1) \rangle
  \nonumber \\
  & = & \frac{2 J+1}{3}\, 1.8 \times 10^{-3}\, {\rm GeV}^5
\label{element3PJ}
\ .
\eeqa

The total $\chi_{\c 1,2}$ cross sections are dominated by colour-octet
processes, as shown in Table~\ref{table:integrated}. 
The (infrared-finite) colour-singlet parts are two orders 
of magnitude smaller than the $\JP$ cross sections. Adding in the 
colour-octet contributions leads to a marked increase. Still, 
the rates are only about $1/20$--$1/10$ of the $\JP$ production rate. 
This confirms that $\chi_{\c J}$ production in $\ee$ annihilation is 
indeed very sensitive to the power counting of long-distance effects 
in quarkonium formation. The non-observation of $\chi_{\c}$'s at CLEO 
\cite{Wolf} (for integrated luminosities of about $3.1\,$fb$^{-1}$ on 
the $\Upsilon(4S)$ and about $1.6\,$fb$^{-1}$ off the resonance)
already causes problems \cite{photo-ratio} for the CEM, where 
the $\chi_{\c J} : \JP$ ratio is predicted to be the same
as was measured in fixed-target \cite{CERN7170} and collider
\cite{collider-ratio} hadroproduction, \ie\ about $1$ ($5/3$) for 
$J=1$ ($J=2$). 
 
The rates in Table~\ref{table:integrated} are for direct production,
\ie\ excluding the production of $\chi_{\c J}$ in the decay of other
hadrons. $\B$-meson decay contributions can be removed experimentally, 
e.g.\ by a cut on $z$ at CLEO. There is also a contribution from 
$\psi(2S)$ decays, $\psi(2S) \rightarrow \gamma + \chi_{\c J}$. 
We estimate the $\chi_{\c 1}$ cross sections to be 
$11\,$fb for process (\ref{cc13S1cc}), $19\,$fb
for (\ref{cc13S1gg}), $8.3\,$fb for (\ref{cc81S0g}), $19\,$fb for
(\ref{cc83PJg}), and $0.64\,$fb for (\ref{cc83S1qq}). 
In total we hence expect about $60\,$fb ($10$\% less for $\chi_{\c 2}$), 
with about one half due to colour-octet mechanisms. We conclude 
that direct $\chi_{\c J}$ production dominates over indirect one.

\begin{figure}
\psfig{file=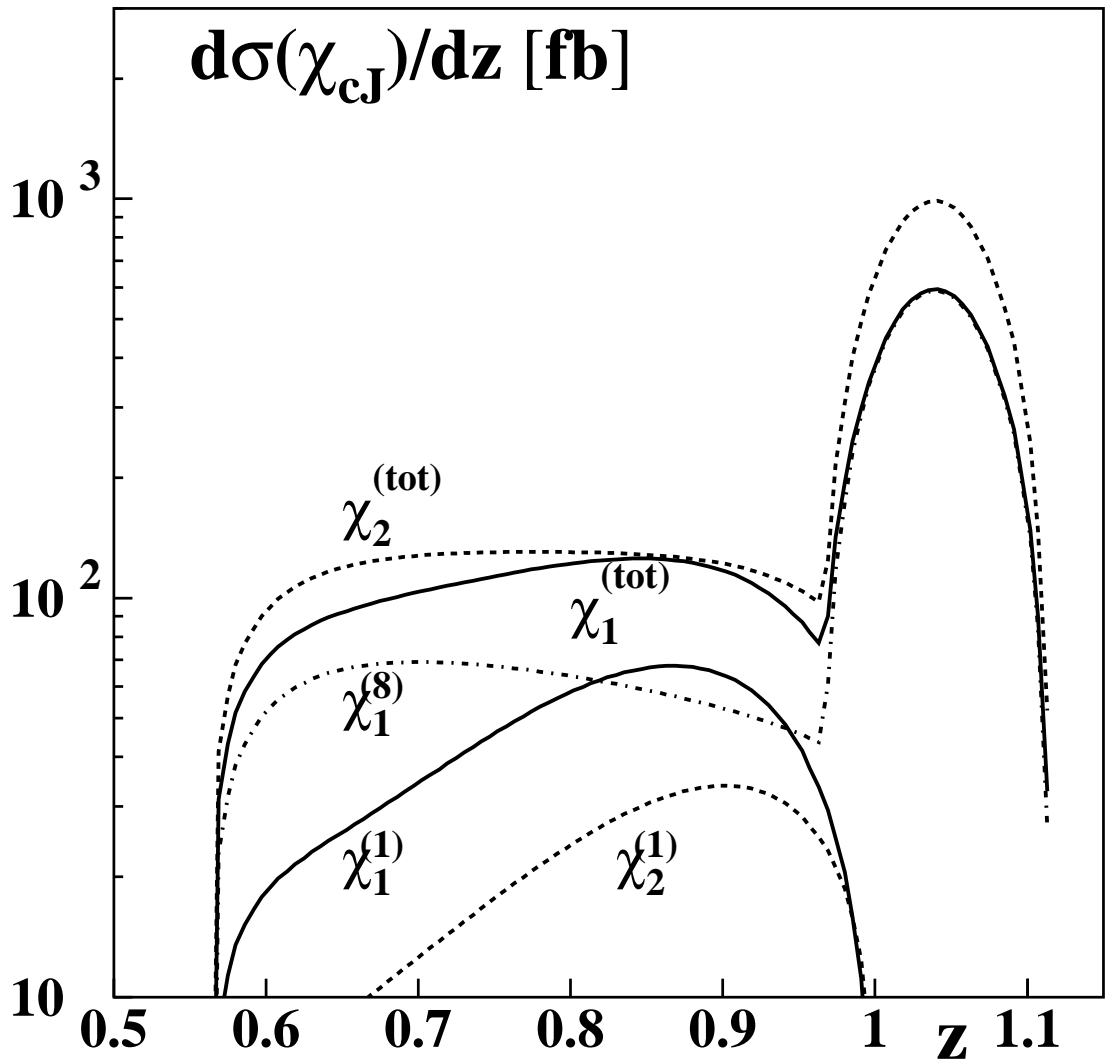,width=\textwidth}
\caption{Energy distribution of $\sigma(\ee \rightarrow \chi_{\c 1,2} + X)$ 
at \protect\ecm: total, colour-singlet, and colour-octet $\chi_{\c 1}$ 
contributions, and total and colour-singlet $\chi_{\c 2}$ 
contributions. The peak from the  $\ee \to \c\cbar_8(^1S_0, \, ^3P_J) + \g$
processes at large $z$ has been smeared over a range of width
$\sim v^2$. In a physical process such a smearing follows from
the energy transfer in the non-perturbative transition.\label{fig:energy}}
\end{figure}

Figure~\ref{fig:energy} shows the energy distribution of
direct $\chi_{\c J}$ production. 
A steep rise of the distribution at large $z$
signals the importance of the $\ee \to \c\cbar_8(^3P_J) + \g$
process. Signatures for the other colour-octet mechanisms are 
also clearly visible, e.g.\ a $\chi_{\c 1} : \chi_{\c 2}$ ratio 
close to $1$ at $z<1$, as opposed to the ratio $< 0.5$ obtained 
from colour-singlet processes alone.

\begin{figure}
\psfig{file=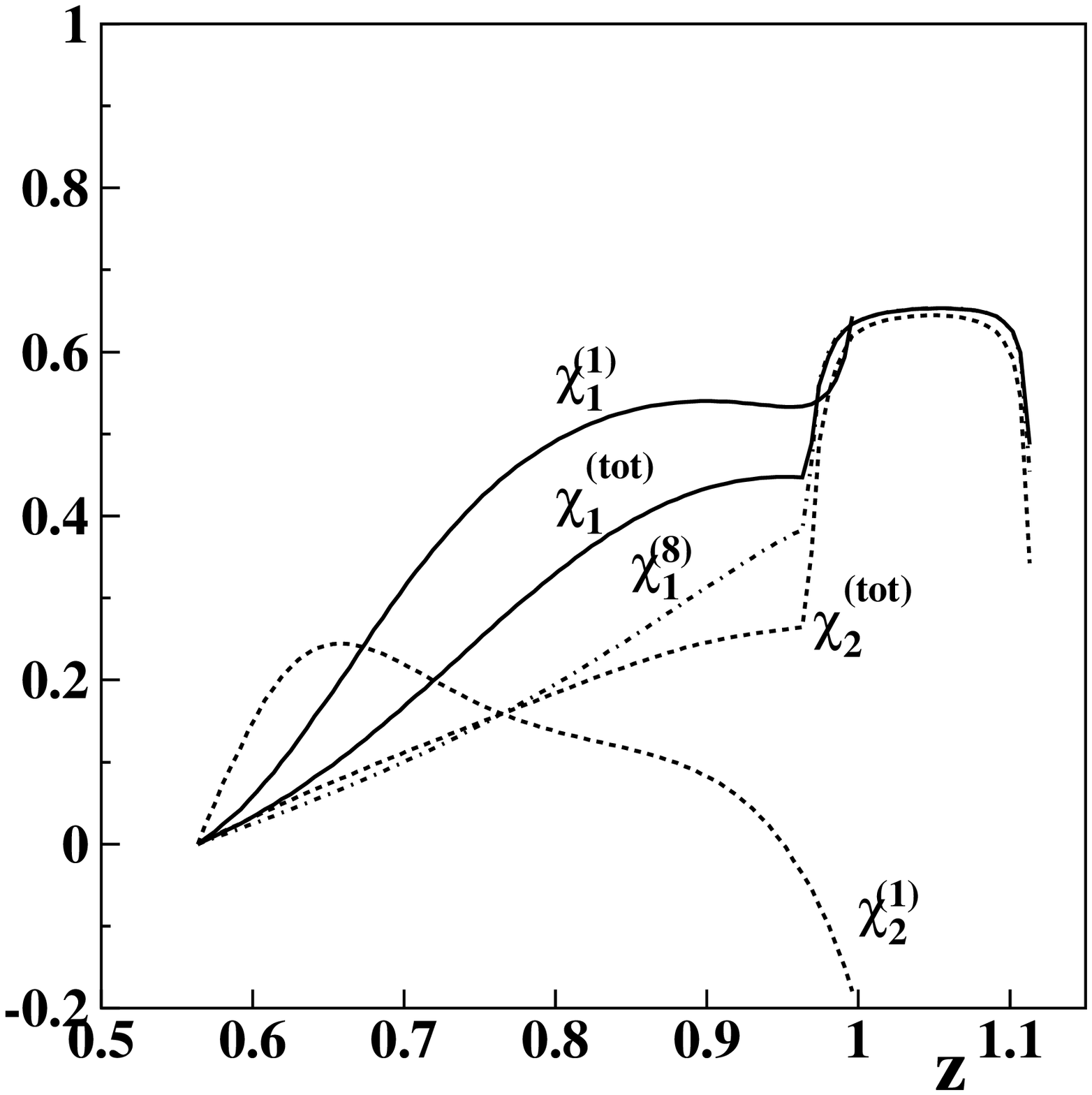,width=\textwidth}
\caption{Energy dependence of the angular coefficient $\alpha$ in
$\ee \rightarrow \chi_{\c 1,2} + X$ at \protect\ecm: 
total, colour-singlet, and colour-octet $\chi_{\c 1}$ contributions, and
total and colour-singlet $\chi_{\c 2}$ contributions. \label{fig:angular}}
\end{figure}

With a separate measurement of $\chi_{\c 1,2}$ and with sufficient
statistics to determine the double-differential cross section
$\d^2\sigma/\d z \, \d\cos\theta$, the angular distribution parameter
$\alpha(z)$ will also serve to identify NRQCD production channels.
As shown in Fig.~\ref{fig:angular}, the colour-octet value of $\alpha(z)$ 
for $\chi_{\c 1}$ is significantly lower than the
colour-singlet value, whereas for $\chi_{\c 2}$ the colour-octet
value (at large $z$) is significantly higher than the colour-singlet one.
 
In summary, inclusive $\chi_{\c J}$ production is a powerful tool
to establish the size of higher Fock state contributions in the
formation of heavy bound states. 
Different theoretical approaches rely on different power-counting rules 
resulting in markedly different cross-section ratios. 
For example, the $\chi_{\c 2}:\JP$ ratio is as low as $1/100$ if 
only $\c\cbar$ pairs in the leading Fock state contribute, while 
it is of order $1$ if bound-state formation proceeds dominantly 
through (colour-singlet or -octet) $S$-wave $\c\cbar$ pairs, 
as it does in the CEM. The velocity-scaling rules of NRQCD 
yield a ratio of about $1/12$, since 
the presence of colour-octet
processes induced by higher Fock states is much more pronounced for
$\chi_{\c J}$ production than for $\JP$ production. This is partly
because colour-octet processes enter $\chi_{\c J}$ production at the
same order as the colour-singlet ones, and partly because $C$-parity
suppresses the process $\ee \rightarrow \c\cbar \g\g$, which dominates
$\JP$ production. 

Signatures for colour-octet contributions do not only 
show up in a dramatic increase in total $\chi_{\c J}$ production rates. 
Differences are also clearly visible in the energy and angular 
distributions. The expected rates could already be visible at CLEO
with current statistics, and will definitely 
be measured in the near future at CLEO and at $\B$-factories.

M.V.\ wishes to acknowledge financial support from 
Suomalainen Tiedeakatemia, V\"ais\"al\"an rahasto.

\begin{table}
\caption{
Integrated cross sections and angular coefficient $\alpha$ 
for $\ee \rightarrow \chi_{\c 1,2}+ X$ at \protect\ecm. The colour-octet
$\chi_{\c 2}$ cross sections are a factor $5/3$ larger than the corresponding 
$\chi_{\c 1}$ ones. The cut $z>0.693$ serves to exclude $\chi_{\c J}$'s
originating from $\B$-meson decays. \label{table:integrated}}
\begin{tabular}{lrrrr}
 & \multicolumn{2}{c}{All $z$}
 & \multicolumn{2}{c}{$z>$ 0.693} \\ 
 & \multicolumn{1}{c}{$\sigma$ [fb]}
   & \multicolumn{1}{c}{$\alpha$}
   & \multicolumn{1}{c}{$\sigma$ [fb]}
   & \multicolumn{1}{c}{$\alpha$}
 \\ \tableline
  $\c\cbar + \c\cbar_1 (^3P_1 ) \rightarrow \chi_{\c 1}$    
 &      $18.1$ &    $0.44$
 &      $15.3$ &    $0.49$
  \\ 
  $\c\cbar + \c\cbar_1 (^3P_2 ) \rightarrow \chi_{\c 2}$  
 &      $ 8.4$ &    $0.10$
 &      $ 7.5$ &    $0.09$
  \\ 
  $\c\cbar + \c\cbar_8 (^3S_1 ) \rightarrow \chi_{\c 1}$ 
 &      $ 6.1$ &    $0.24$
 &      $ 4.0$ &    $0.35$
  \\ 
  $\q\qbar + \c\cbar_8 (^3S_1 ) \rightarrow \chi_{\c 1}$  
 &      $15.6$ &    $0.26$
 &      $11.7$ &    $0.34$
  \\ 
  $\g\g     + \c\cbar_8 (^3S_1 ) \rightarrow \chi_{\c 1}$  
 &      $ 5.5$ &    $-0.03$
 &      $ 4.2$ &    $-0.05$
  \\ 
  $\g        + \c\cbar_8 (^1S_0 ) \rightarrow \chi_{\c 1}$    
 &      $ 4.1$ &    $1.00$
 &      $ 4.1$ &    $1.00$
  \\ 
  $\g        + \c\cbar_8 (^3P_J ) \rightarrow \chi_{\c 1}$  
 &      $45.6$ &    $0.64$
 &      $45.6$ &    $0.64$
  \\ 
 Total $\chi_{\c 1}$
 &      $95.0$ &    $0.47$
 &      $84.8$ &    $0.53$
  \\ 
 Total $\chi_{\c 2}$
 &     $136.5$ &    $0.46$
 &     $123.4$ &    $0.51$
  \\ 
\end{tabular}
\end{table}

\begin{references}
\bibitem{notation-footnote}
We use the spectroscopic notation for angular momenta, and the
subscript $1$ $(8)$ denotes a colour-singlet (colour-octet) state.
\end{references}
\end{document}